# Mechanochemical synthesis of pnictide compounds and superconducting $Ba_{0.6}K_{0.4}Fe_2As_2$ bulks with high critical current density

J D Weiss, J Jiang, A A Polyanskii and E E Hellstrom

Applied Superconductivity Center, National High Magnetic Field Laboratory, Florida State University, FL 32310, USA

E-mail: Weiss@magnet.fsu.edu

**Abstract.** $BaFe_2As_2$ (Ba-122) and $(Ba_{0.6}K_{0.4})Fe_2As_2$ (K-doped Ba-122) powders were successfully synthesized from the elements using a reaction method, which incorporates a mechanochemical reaction using high-impact ball milling. Mechanically-activated, self-sustaining reactions (MSR) were observed while milling the elements together to form these compounds. After the MSR, the Ba-122 phase had formed, the powder had an average grain size < 1 μm, and the material was effectively mixed. X-ray diffraction confirmed Ba-122 was the primary phase present after milling. Heat treatment of the K-doped MSR powder at high temperature and pressure yielded dense samples with high phase purity but only granular current flow could be visualized by magneto optical imaging. In contrast, a short, low temperature, heat treatment at ambient pressure resulted in global current flow throughout the bulk sample even though the density was lower and impurity phases were more prevalent. An optimized heat treatment involving a two-step, low temperature, heat treatment of the MSR powder produced bulk material with very high critical current density above 0.1 $MAcm^{-2}$ (4.2 K, 0 T).

**Introduction**

Since their discovery, the ferropnictide materials have spawned great interest in the physics and material science communities for investigating the mechanism of superconductivity. They also show promise for applications [1-3]. Many superconducting devices require long wires or tapes of Mechanochemical synthesis of pnictide compounds and superconducting Ba$_{0.6}$K$_{0.4}$Fe$_2$As$_2$ bulks with high critical current density homogeneous superconducting material with well-connected grains that allow supercurrent to flow through the bulk material. Most of the ferropnictide wires and bulks made to date exhibit extrinsic weak link behavior as a result of low density, cracking, and impurity phases that wet the grain boundaries [4–11] in addition to possible intrinsic weak link behavior attributed to grain boundaries [12,13]. Figure 1 is an SEM image of a Ba-122 bulk prepared by a typical grind-and-react route that shows clearly some of the impurity phases that often wet grain boundaries and block current. Safe and efficient reaction methods are needed to produce phase-pure bulk material that can be used to make superconducting devices.

The risk of an explosion from the high vapor pressure of As at high temperature and its being poisonous have resulted in reaction methods for ferropnictides that typically involve slow heating, intermediate grinding steps, and/or high pressure synthesis [4,6,7,11,14–19]. Multistep synthesis routes that use binary compounds, such as M-As (M=Ba, Sm, K, Co, Fe, La, or Nd) instead of the elements to prevent the vaporization of As are also common [20–22], but synthesizing the M-As compounds has the same risks. The origin of the present study was our observation that occasionally when hand grinding the elements to make Ba-122 or M-As binary compounds in a mortar and pestle, exothermic, self-sustaining reactions (SSR) occurred that were activated by heat from friction between the mortar and pestle. These unintentional reactions further increase the hazards and difficulties of synthesizing these materials because the reactions have large heats of formation that heat the powders, which vaporizes As and other constituents.

Igniting an SSR with a mechanical impact, such as with the mortar and pestle, is referred to as a mechanically-activated self-sustaining reaction (MSR). We realized we could use high-energy ball milling to create controlled MSRs. We initiated MSR reactions in mixtures of Ba and As, and Sm and As that we subsequently used to synthesize Ba-122 or SmFeAsO$_{1-x}$F$_x$ (Sm-1111). We advanced from these binary systems and investigated whether we could use MSR reactions to form Ba-122 directly from the elements. Direct synthesis of Ba-122 by MSR is successful and is reported here. The MSR powders have the desired overall composition, i.e., there is no loss of volatile elements, and the particles have submicron grain sizes that aid in homogeneously mixing the constituents. Having a well-mixed powder with very fine grain size is desirable because it promotes the formation of more homogeneous, phase-pure material. Using MSR powders can help retain more volatile elements such as As and K in the final product since it allows for shorter reaction times at lower temperatures than are typically required when using standard solid-state synthesis [23,24]. In this study we found that the MSR occurred just minutes

Mechanochemical synthesis of pnictide compounds and superconducting $Ba_{0.6}K_{0.4}Fe_2As_2$ bulks with high critical current density after starting the milling. The MSR-produced powder was characterized before and after the MSR reaction, and after different heat treatments to form bulk samples of K-doped Ba-122.

**Experimental details**

All handling of material and milling was carried out in Ar-atmosphere glove boxes. We used As particles (20 mesh, Alfa Aesar), Ba pieces (~20 mesh, Alfa Aesar), Fe powder (200 mesh, Alfa Aesar) and K chunks (Alfa Aesar) in our experiments. Typically 7 g batches of Ba-122 powder with nominal composition $BaFe_2As_2$ and $(Ba_{0.6}K_{0.4})Fe_2As_2$ were synthesized by combining stoichiometric amounts of Ba, Fe, As, and K in a zirconia milling vial (SPEX 6.4 cm diameter, 6.8 cm long) with two 12.7 mm diameter zirconia milling balls. Samples were milled using a high-impact milling machine (SPEX 8000M) for 1 hour inside the Ar glove box. To monitor the exothermic MSR reactions, thermocouples were attached to the surface of a smaller steel milling jar (3.8 cm diameter, 6.6 cm long) loaded with three 9.5 mm diameter steel milling balls and a 3 g batch of the elements. Temperature as a function of milling time was measured for reactions between Ba and As, and the elements to form Ba-122 and K-doped Ba-122.

After 60 min of milling in the zirconia jar, the powder was removed, and then wrapped with Nb foil and placed into stainless steel ampoules. The ampoules were evacuated, welded shut, compressed with a cold isostatic press (CIP) (AIP Columbus, OH) at 275 MPa to press the powder into a pellet, and then heat treated. Three different heat treatments were used and the amount of powder used to make the pellet was adjusted for the specific heat treatment. A sample with 2 g of powder was heat treated for 12 hours at 1120°C then ramped down at 4 °C/h and held for 20 hours at 900 °C in a hot isostatic press (HIP) (AIP, Columbus, OH) at 193 MPa (sample name = 1120 °C HIP). A sample made with 0.5 g of powder was heated for 10 hours at 600 °C at ambient pressure (1 bar) (sample name = 600 °C AP). The third sample was made by heat treating 5 g of the powder in the HIP under 192 MPa of pressure at 600 °C for 20 hours. This sample was then re-milled and 1.5 g of powder was heat treated again in the HIP (600 °C, 193 MPa) for 10 hours (sample name = 600 °C HIP). The baseline undoped Ba-122 sample shown in figure 1 was made using our previous technique of hand grinding stoichiometric amounts of As, Fe, and $Ba_3As_2$ in a zirconia mortar and pestle for 60 min. No MSR was observed for this sample. This baseline Ba-122 sample was placed in Nb foil, encapsulated, CIPped, and then HIPped as described above for the sample heat treated at 1120 °C.

Mechanochemical synthesis of pnictide compounds and superconducting $Ba_{0.6}K_{0.4}Fe_2As_2$ bulks with high critical current density

XRD patterns of the milled powder and heat treated samples were recorded by a HUBER imaging plate Guinier camera (model 670) using Cu K$\alpha_1$ radiation with a Ge monochrometer in steps of 0.005°. Generic transparent tape (Office Max) was used to hold the powders during XRD to prevent sample contamination. Magnetization of the heat-treated samples was measured by a SQUID magnetometer (Quantum Design: MPMS-XL5s) and a 14 T Oxford vibrating sample magnetometer (VSM) with the magnetic field parallel to the sample's length. Magneto optical (MO) imaging with a 5 μm thick Bi-doped garnet indicator film placed onto the sample surface was used to image the normal field component produced by magnetization currents induced by magnetic fields applied perpendicular to the sample's surface. The SEM samples were dry polished using SiC paper up to 800 grit and then wet polished with a 0.05 μm $Al_2O_3$ - isopropyl solution for 30 minutes. SEM images were taken with a Zeiss SEM at 20 kV (Model 1540).

**Results**

Figure 2 shows the vial temperature as a function of milling time for powders that have BaAs, $BaFe_2As_2$ and $Ba_{0.6}K_{0.4}Fe_2As_2$ stoichiometries. The MSR was observed as the sharp increase in vial temperature. The baseline shows a steady temperature increase due to energy input from impact of the milling balls and due to the ambient temperature rise within the glove box from the SPEX mill components such as the motor. The XRD patterns of the K-doped Ba-122 powder just before the MSR and after 60 minutes of milling are displayed in figure 3(a) and 3(b), respectively. An SEM image of the milled powder is included in figure 3(b) showing a submicron particle size. The XRD pattern (figure 3(b)) shows that K-doped Ba-122 is the primary crystalline phase indicating that K-doped $BaFe_2As_2$ formed during the MSR reaction. There are minor extra peaks in the pattern due to the tape we used to hold and cover the sample. As shown by XRD (left) and SEM (right) in figure 3(c), heating at 1120 °C turned the MSR powder into nearly pure K-doped Ba-122 bulk with grain size over 60 μm. A small amount of Fe+$Fe_2As$ (<1 vol%) was identified by energy-dispersive X-ray spectroscopy (EDS). Figures 3(d) and 3(e) show the XRD patterns and SEM micrographs of samples 600 °C AP and 600 °C HIP, respectively. The XRD patterns show sharp diffraction peaks and only trace minority phase peaks from the FeAs phase. SEM revealed that both samples contained more voids and FeAs phase than observed in the sample heat treated at 1120 °C, but the FeAs existed in the 600 °C samples as discrete particles that did not wet the grain boundaries and accounted for about 2-3 vol% of the sample.

Mechanochemical synthesis of pnictide compounds and superconducting $Ba_{0.6}K_{0.4}Fe_2As_2$ bulks with high critical current density

The volumetric susceptibility measurements, shown in figure 4, were taken after zero-field-cooling (ZFC) the samples to 5 K and then applying a magnetic field of 20 Oe. All three samples show a sharp transition and nearly perfect diamagnetism. Sample 1120 °C HIP has the sharpest transition, indicating good electromagnetic homogeneity throughout the superconducting bulk, but a slight residual magnetization above $T_c$ from the magnetic impurity phase observed by SEM. $T_c$ decreases slightly from 38.5 K for sample 1120 °C HIP to around 37 K for the samples heat treated at 600 °C (see table 1). Sample 600 °C HIP has a sharper transition than sample 600 °C AP, indicating the two-step, high-pressure synthesis improved the electromagnetic homogeneity.

MO imaging was used to visualize the local field profile produced by magnetization currents induced by an external magnetic field applied perpendicular to each sample's surface after zero field cooled (ZFC) and by the remnant magnetization when the field was removed. The MO image in figure 5(a) of sample 1120 °C HIP shows flux penetration at the grain boundaries and local regions of partial flux penetration within individual grains produced by granular current flow, but no indication of flux produced by current flowing over the whole sample. Figure 5(b) shows the corresponding remnant field. Figures 5(c) and 5(d) are MO images of sample 600 °C AP showing a nearly uniform rooftop pattern of magnetic flux density produced by bulk current flow throughout the entire bulk sample after an external magnetic field is applied and then removed, respectively. Figure 5(e) shows that sample 600 °C HIP has a much more uniform rooftop pattern of magnetic flux density and less flux penetration than sample 600 °C AP, indicative of better connectivity. Figure 5(f) is the corresponding remnant field produced by the bulk when the field was removed. Figure 6 shows the magnetization $J_c^{mag}$ as a function of applied magnetic field for both 600 °C samples. $J_c$ was calculated using the Bean model, using the sample's bulk dimensions, from magnetization hysteresis loops taken at several temperatures. The high $J_c$, which is above 0.1 $MAcm^{-2}$ (<10 K, SF) for sample 600 °C HIP, is consistent with the highest values reported for ferropnictide wires [2].

**Discussion**

The literature shows that the starting particle size of the powders being milled affects the time to ignition and the kinetics of the combustive reaction front [25]. Typical MSR reactions reported in the literature begin with fine powders (>100 mesh) and the time to ignition can vary from minutes to hours

Mechanochemical synthesis of pnictide compounds and superconducting Ba$_{0.6}$K$_{0.4}$Fe$_2$As$_2$ bulks with high critical current density depending on the reactants and milling conditions [23,25,26]. The temperature at which ignition occurs for an SSR is thought to decrease with decreasing particle size and with increasing mixing of these smaller particles [25]. After milling for some time, the ignition temperature is decreased until it is lower than the local temperature generated by the impact between the milling vial and media, and then the MSR occurs. In addition, milling causes a general increase in powder temperature and creates a high density of defects in the powder that can affect the reaction kinetics of the MSR reaction. The sharp increase in the temperature of the outer surface of the milling vial within 1 to 10 min indicated an exothermic MSR reaction had occurred in doped and undoped Ba-122 samples as well as BaAs in a much shorter time than reported in the literature for most other MSRs [23,25,26].

Ba and K are particularly difficult to work with when grinding powders together for normal solid-state reactions. They are difficult to grind into high-purity, fine-grain size powder because they are both soft and easily contaminated with oxygen. Using MSR alleviates these problems because the MSR, which occurs rapidly using high-impact ball milling, produce a brittle, less reactive intermetallic compound that mixes better during the ball milling. The MSR produced powder also has finer grain size than can be obtained with hand grinding, so the MSR powder sinters at lower temperatures. We have studied MSR reactions between other electropositive metals and pnictide reactants that are not reported here and based on these results we believe that MSR reactions will occur in other systems, including superconducting ferropnictides, that combine highly electropositive elements like Ba, Sr, K, Sm, Nd with electronegative elements such as As, P, and O. Here too, the MSR reaction should lead to higher phase purity.

The heat treatments were chosen to cover a wide range of processing conditions. The high-pressure and high-temperature reaction used for sample 1120 °C HIP resulted in a material with very high phase purity (< 1 vol% impurities observed by SEM) and very high density (~ 98% dense). This heating schedule was previously developed and optimized by our group for making high quality Ba-122 samples from the elements without using MSR powder. The high $T_c$ and sharp transition indicate the superconducting part of the material is uniform in composition. However, no bulk current flow was observed in this sample because of cracks and the segregation of impurity phases to the grain boundaries, a common problem for bulks processed at high temperatures [7,8,11,19].

The low reaction temperature (600 °C) was chosen to prevent FeAs liquid phases from forming and to stay below the vaporization temperature of As. Staying below the liquidus line in the Fe-As phase diagram insures these impurities won't wet the grain boundaries and staying below the vaporization

Mechanochemical synthesis of pnictide compounds and superconducting $Ba_{0.6}K_{0.4}Fe_2As_2$ bulks with high critical current density temperature of As keeps residual As from subliming, which could cause an explosion hazard and risk of As escaping the reaction crucible. The lower temperature also results in a very small grain size that yields fewer cracks and may improve the intergranular current density by improving the flux vortex dynamics as speculated elsewhere [2]. We see by MO imaging and magnetization measurements that even though sample 600 °C AP is less dense (~68% theoretical density) and has a broader temperature dependence of $\chi^{vol}$ than sample 1120 °C HIP, which can be attributed to an inhomogeneous distribution of potassium, there is significant intergranular current transport. It is interesting to note that that the 68% theoretical density (table 1) is close to the theoretical density (74 %) obtained by packing monosized, spherical particles together. For sample 600 °C HIP, we used the second milling and second heat treatment to improve homogeneity and HIP processing to increase the density. The improved electromagnetic properties compared to sample 600 °C AP can be seen by the sharper $\chi^{vol}$ transition and the density was increased to ~92% of the theoretical density.

The MO images in figure 5 clearly show that cracks and grain boundary phases can significantly block current for samples 1120 °C HIP and 600 °C AP. Sample 600 °C HIP was free of cracks and grain-wetting phases and had the highest $J_c$ as shown in figure 6 (b) with weak field dependence at high fields and temperatures. Its $J_c$ is above $10^5$ $Acm^{-2}$ at low fields at and below 10 K, which is around the values required for applications. If $J_c$ can be improved further by a factor of 10 in field, K-doped Ba-122 would be useful for high field applications in the temperature range obtainable by cryo-coolers and liquid $H_2$.

**Conclusions**

MSR reactions that can occur when hand grinding elements to form Ba-122 can be hazardous. We studied controlled MSR reactions to form Ba-122 and K-doped Ba-122 using high-impact ball milling. Pre-reacting mixtures of the elements by MSR yields submicron size as-milled powder with Ba-122 as the primary crystalline phase. After a high temperature HIP heat treatment (1120 °C), the K-doped Ba-122 bulk materials made from MSR powder have high density (~98.5% dense) and higher phase purity (< 1% secondary phase by SEM) than can be obtained by HIPping hand-ground powders, but they did not carry bulk supercurrent due to cracks and the impurity phase wetting, and blocking, the grain boundaries. A low temperature reaction at ambient pressure and 600 °C, which is well below the melting point of the FeAs impurity phase, resulted in bulk material that could carry over 10 $kAcm^{-2}$ (4.2 K, SF)

Mechanochemical synthesis of pnictide compounds and superconducting Ba$_{0.6}$K$_{0.4}$Fe$_2$As$_2$ bulks with high critical current density despite being less dense (~ 68%) and having more impurity phases (>1 vol%) than the material made by HIPping the MSR powder at 1120 °C. Using a two-step HIP heat treatment at 600 °C resulted in a material that was ~92% dense, with 1-3 vol% impurity phases and good connectivity, carrying over $10^5$ Acm$^{-2}$ (<10 K, SF). We believe our new MSR synthesis route can be used to safely synthesize other ferropnictides and will result in bulk materials that have higher phase purity with fewer reaction steps than traditional reaction pathways.

## Acknowledgments

We would like to thank W L Starch and V S Griffin for technical support. This work is supported by NSF DMR-1006584 and by the National High Magnetic Field Laboratory, which is supported by the National Science Foundation under NSF DMR-1157490 and by the State of Florida.

Mechanochemical synthesis of pnictide compounds and superconducting Ba$_{0.6}$K$_{0.4}$Fe$_2$As$_2$ bulks with high critical current density

**Table 1.** Material properties of superconducting bulks.

| Sample | $T_c$ (K) | $J_c$(SF, 4.2 K) (kAcm$^{-2}$) | $J_c$(10T, 4.2 K) (kAcm$^{-2}$) | Density (g/cm$^3$) | Fraction of theoretical density |
|---|---|---|---|---|---|
| **1120 °C HIP** | 38.7 | NA | NA | 5.76 (+-0.04) | 0.985 (+- 0.007) |
| **600 °C AP** | 37.7 | 11.0 | 0.9 | 4.0 (+- 0.1) | 0.68 (+- 0.02) |
| **600 °C HIP** | 37.1 | 125.0 | 10.4 | 5.4 (+- 0.1) | 0.92 (+- 0.02) |

Mechanochemical synthesis of pnictide compounds and superconducting $Ba_{0.6}K_{0.4}Fe_2As_2$ bulks with high critical current density

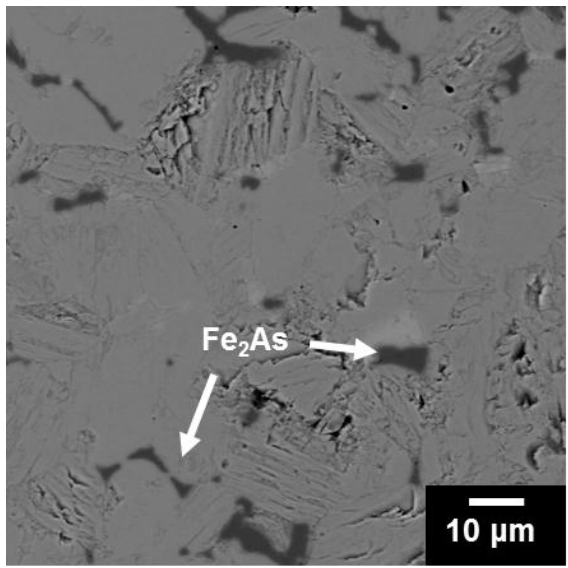

Figure 1 – SEM image of $BaFe_2As_2$ made by hand grinding Fe, $Ba_3As_2$, and As followed by HIP treatment at 1120 °C. The image shows voids and $Fe_2As$ that wets grain boundaries.

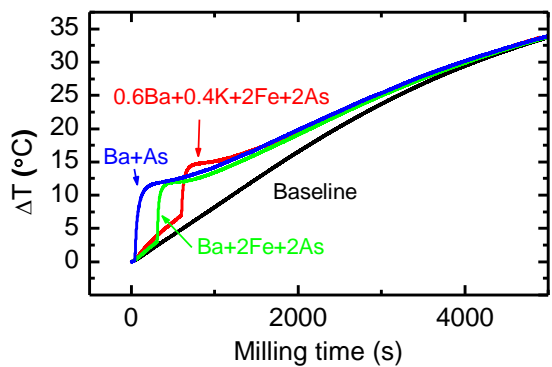

Figure 2 – Milling vial temperature as a function of milling time for three different MSR reactions. The baseline was obtained by milling fully reacted and milled $BaFe_2As_2$ powder in which an MSR was not observed.

Mechanochemical synthesis of pnictide compounds and superconducting $Ba_{0.6}K_{0.4}Fe_2As_2$ bulks with high critical current density

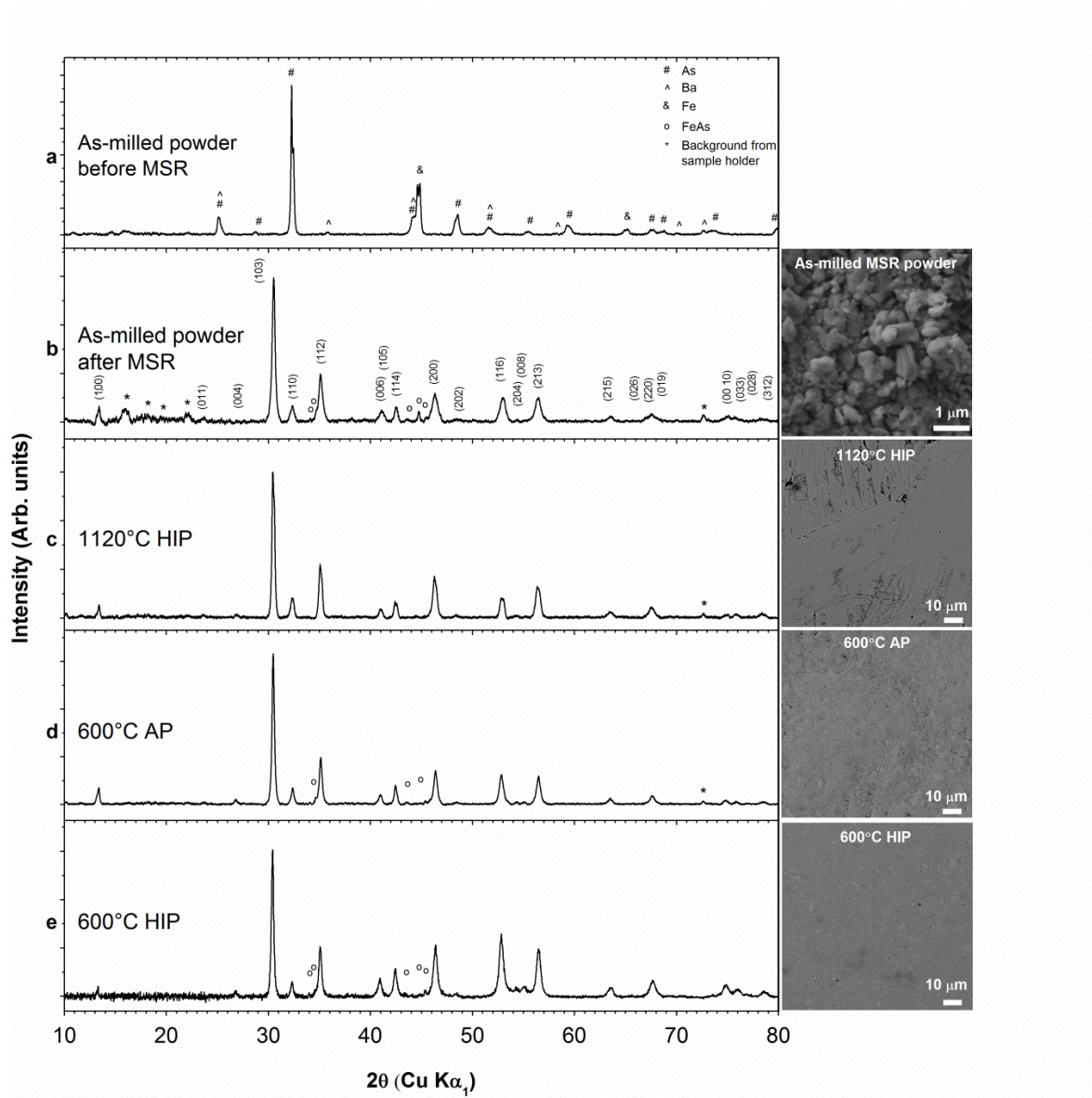

Figure 3 – (a) XRD pattern of milled 0.6 Ba + 0.4 K + 2 Fe + 2 As powder before MSR. XRD pattern and SEM image of (b) milled powder after MSR when the $Ba_{0.6}K_{0.4}Fe_2As_2$ phase had formed, (c) after the 1120 °C HIP heat treatment of MSR powder, (d) after the 600 °C AP heat treatment of MSR powder, and (e) after the 600 °C HIP heat treatment of the MSR powder.

Mechanochemical synthesis of pnictide compounds and superconducting Ba$_{0.6}$K$_{0.4}$Fe$_2$As$_2$ bulks with high critical current density

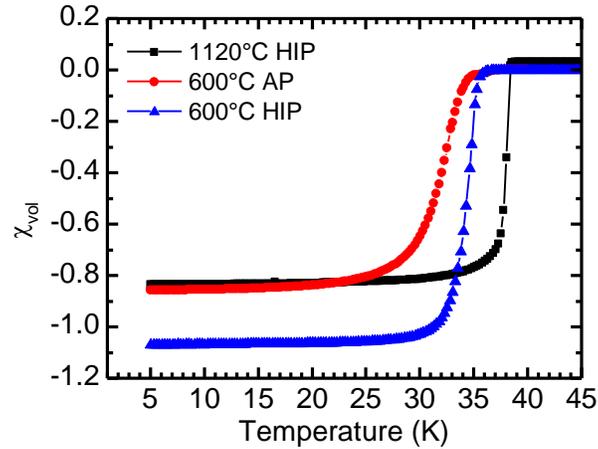

Figure 4 – Temperature dependence of the volumetric susceptibility under zero field cooling (ZFC) in an external field of 20 Oe for (Ba$_{0.6}$K$_{0.4}$)Fe$_2$As$_2$ samples 1120 ° C HIP, 600 ° C AP, and 600 ° C HIP.

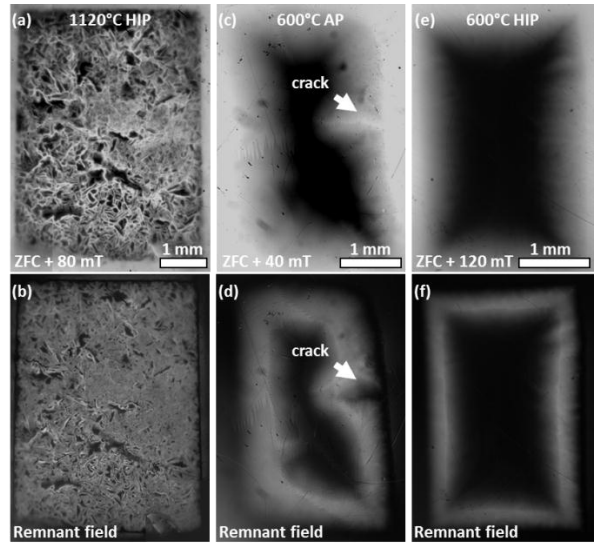

Figure 5 – Magneto optical images showing flux penetration after zero-field-cooling (ZFC) the sample to 10 K and applying the magnetic field shown in the image, for samples (a) 1120 ° C HIP, (c) 600 ° C AP, and (e) 600 ° C HIP. (b), (d), and (f) are magneto optical images of the remnant magnetic flux after the field was removed corresponding to (a), (c), and (e), respectively.

Mechanochemical synthesis of pnictide compounds and superconducting $Ba_{0.6}K_{0.4}Fe_2As_2$ bulks with high critical current density

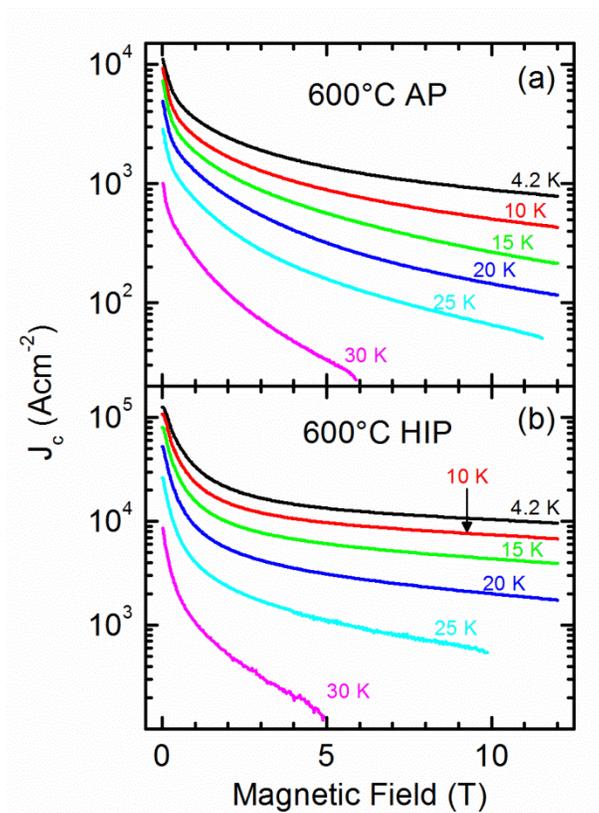

Figure 6 – Magnetic field dependence of the critical current density calculated from magnetization measurements at 4.2 K, 10 K, 15 K, 20 K, 25 K, and 30 K for samples (a) 600 °C AP, and (b) 600 °C HIP.